# Ring Oscillators for Clocking Reversible Superconducting Circuits and Fabrication Process Benchmarking

Vasili K. Semenov and Yuri A. Polyakov[*]
Stony Brook University
Stony Brook, NY 11794, USA
Vasili.Semenov@StonyBrook.edu
Yuri.Polyakov@StonyBrook.edu

Sergey K. Tolpygo
Lincoln Laboratory
Massachusetts Institute of Technology
Lexington, MA 02420, USA
Sergey.Tolpygo@ll.mit.edu

*Abstract*—Existing concepts of reversible superconducting circuits as well as demonstrated adiabatic circuits require three-phase bias/clock signals generated by room temperature sources. A while ago, we suggested that a multi-phase bias/clock could be provided by a local Josephson junction-based generator. The generator looks like a long annular Josephson junction, only composed of discreet elements - junctions and inductors, and closed into a ring via a flux pump to inject any required number of vortices into the ring. A steady motion of the vortices forced by a uniformly distributed dc bias current applied to the ring is accompanied by a nearly harmonic ac currents flowing via the Josephson junctions (JJs) connected in series with small inductors. These ac currents serve as multi-phase bias/clock for nSQUID-based circuitry. To verify this concept and compare the dissipated energy with $k_B T \ln 2$ threshold, we developed a ring composed of 256 unshunted JJs with 20 µA target critical current, $I_c$. We investigated the behavior of the ring oscillator at each vortex count from 0 to 256. The measured critical current of the ring with vortices was about 0.1 µA per one JJ, which can be explained by unavoidable nonuniformity of the ring components and the influence of fluxes frozen near the ring. The corresponding energy dissipation, about $4k_B T$ per passage of one vortex through one JJ, should be reduced further for prospective experiments with reversible circuits. However, obtained *I-V* characteristics could be of interest for scientists working with long Josephson junctions. Superiority of the fabrication process used in this work is demonstrated by the obtained about 200 times reduction of $I_c$ of the ring with vortices with respect to a single comprising JJ, much larger than in any previously described case.

*Keywords— superconductor electronics, reversible computing*

## I. Introduction

Three decades ago, reversible digital circuits were as mysterious and attractive as qubit circuits nowadays [1], [2]. Since then some reversible circuits evolved into adiabatic circuits that successfully compete with energy-efficient SFQ circuit families [3]. Unfortunately, the numerically estimated energy dissipation of modern adiabatic circuits ($\sim 200 k_B T$) is too high in comparison with the thermodynamic threshold energy

$$E_{th} = k_B T \ln 2. \qquad (1)$$

Besides, the measurement of energy dissipation in AC-biased circuits is quite complicated because of the unavoidable energy loss in the clock wiring connecting a room-temperature AC generator with the circuit at LHe temperature. These losses should also be small compared to (1) and this problem needs to be solved for true reversibility.

We believe that an experimental verification of energy dissipation in reversible circuits is fundamentally important. For example, let us imagine that we were able to reduce the energy dissipation to below, say $10^{-4} k_B T$. Such vanishingly low energy dissipation would imply a strong isolation of the device from a thermal bath. Therefore, we can expect that the device would behave as a quantum rather than classical object. Hence, an aggressive reduction of energy dissipation in reversible circuits could be the shortest way to quantum computing.

As the first step, we figured out how to make a DC-biased reversible circuitry [4]. We invented a Josephson junction based DC/AC converter and integrated it with the reversible circuitry. To be useful for reversible circuits, the energy dissipation in the converter should also be below (1). First experiments with the standalone DC/AC converter have been carried out [5]. Since then, the fabrication technology [6] and our design skills have been substantially improved.

## II. DC/AC Converter

Our DC/AC converter contains a ring oscillator composed of 256 Josephson junctions. The junctions are organized into sections shown in Fig. 1. For correct operation of the converter, the ring should be densely packed with Josephson vortices. In this case, magnetic field is uniformly distributed, and the phase difference of the wave functions grows linearly along the ring:

This research is based upon work supported by the Office of the Director of National Intelligence, Intelligence Advanced Research Projects Activity, via Air Force Contract FA872105C0002. The views and conclusions contained herein are those of the authors and should not be interpreted as necessarily representing the official policies or endorsements of the U.S. Government.

$$\varphi(t,m) = (2\pi/\Phi_0)Vt + km, \quad (2)$$

where $V$ is the DC voltage across the ring. Phase shift $k$ is defined by the number of vortices $n$ injected into the loop via a flux pump [5]. Current

$$I_{jm} = I_c \sin(\varphi(t,m)) \quad (3)$$

flowing via $m$-th Josephson junction and its "payload" $L_j$ models one phase of a multi-phase AC clock for nSQUID based reversible circuitry.

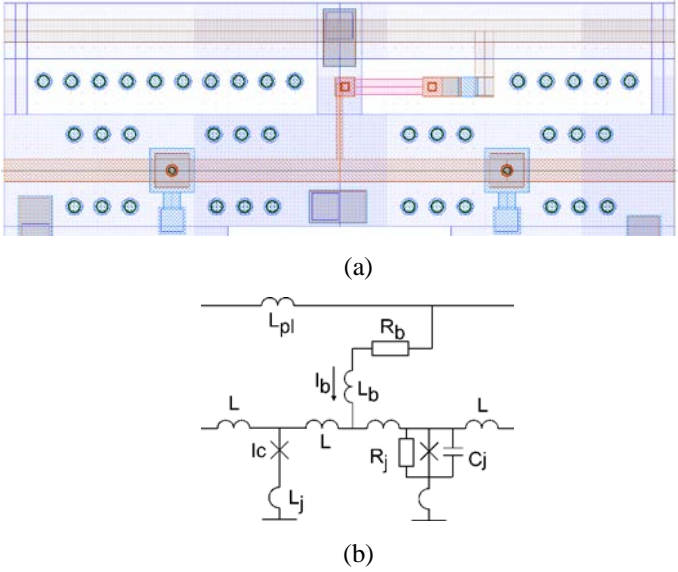

(a)

(b)

Fig. 1. Layout (a) and equivalent circuit (b) of a two-junction section of the ring. The ring target parameters are: L=2.2 pH, inductor area $S_L = 2$ µm x 15 µm = 30 µm$^2$; inductor capacitance $C_L$=10.8 fF, $L_j$=1.09 pH, $I_c$=25 µA, $C_J$=14 fF, $R_J$=800 Ω.

The widths of inductors and the power line are 2 µm; the distance between the individual JJs is 30 µm; the total ring length is about 7.7 mm. Dimensionless inductances are $\beta_L = 0.33$, $\beta_{LJ} = 0.08$. A "digital" flux pump, not shown, allowed us to change the number of injected vortices, $n$. The circuits were fabricated at MIT Lincoln Laboratory using its 4-Nb-layer superconductor electronics fabrication process SFQ3ee.

### III. PROPERTIES OF LONG JOSEPHSON JUNCTION RING

The current-voltage (*I-V*) characteristics of a long circular Josephson junction are shown in Fig. 2. We measured *I-V* characteristics at all unique numbers of injected vortices $n$ corresponding to the first Brillouin zone. In particular, we observed that the curves for $n$ and 256-$n$ vortices practically coincide. Presented experimental results could be of interest to the academic community because our circuit is a discrete-element version of very popular long Josephson junctions; see, e.g., [7]-[9] as a few of most relevant papers among many.

The shown set of current-voltage curves has several unique features. We believe that we have set a record for the range of injected vortices. More importantly, we set a record for the contrast, i.e., the ratio of critical currents without and with injected vortices.

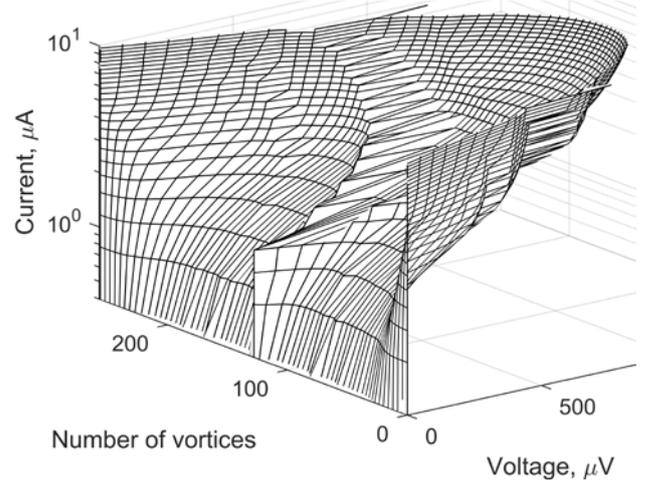

Fig. 2. Current-voltage characteristics of the circuit at various numbers of injected vortices, from 0 to 256.

As we mentioned earlier, the target critical current of one junction was 20 µA. The measured critical current was slightly larger; about 25 µA. Fig. 3 shows the normalized (per one junction) measured critical current of the ring. One induced vortex reduces the normalized critical current down to ~1.1 µA. At larger number of vortices the critical current could be as low as 0.1 µA. So the contrast ranges from 25 to 250. Unfortunately, even the lowest normalized critical current is five times larger than the thermal current corresponding to the thermodynamic threshold (1) $I_{th} = E_{th}/\Phi_0 \approx 0.02$ µA.

### IV. BENCHMARK TEST

The measured critical currents and energy dissipation (per $2\pi$-phase rotation per junction) of the DC/AC converter were still above the thermodynamic threshold. This could be a result of nonuniformities of the circuit components, e.g., spread of their parameters causing pinning of Josephson vortices [8], [9].

Normalized critical currents shown in Fig. 3 provide information about the spread of critical currents. Ideal uniform long Josephson junctions should have zero critical currents, i.e., zero static friction for vortex motion. The discreteness of the circuit is the first factor that should be taken into account. Our analytical [10] and numerical [5] calculations showed that this factor accurately explains the spikes at one vortex per Josephson junction (see Fig. 2) and per two Josephson junctions (see Figs. 2 and 3). However, this discreteness alone is not sufficient to accurately match other data shown in Fig. 3. A much better match has been achieved when a random normally distributed spread of critical currents of the Josephson junctions has been included into our numerical calculations. The best fit to the data in Fig. 3 has been achieved at the simulation with 3.5% or 0.88 µA 1σ spread of critical currents. This spread agrees reasonably well with other spread figures observed for the same fabrication process [6].

A straightforward interpretation of numerical simulations of circuits with randomly varying parameters could be quite confusing. For example, simulated critical currents of the circuit for one set of random critical currents of Josephson junctions could be 2× larger or be only a half of the critical currents of the circuit for another set of critical currents of Josephson junctions. Repeating numerical simulations with different sets of random critical currents of Josephson junctions is required to estimate the range of critical currents of the ring. We think that it could be possible to find a proper set of critical currents of Josephson junction that would exactly match all measured critical currents shown in Fig. 3. However, such "back-engineering" would be a time consuming task.

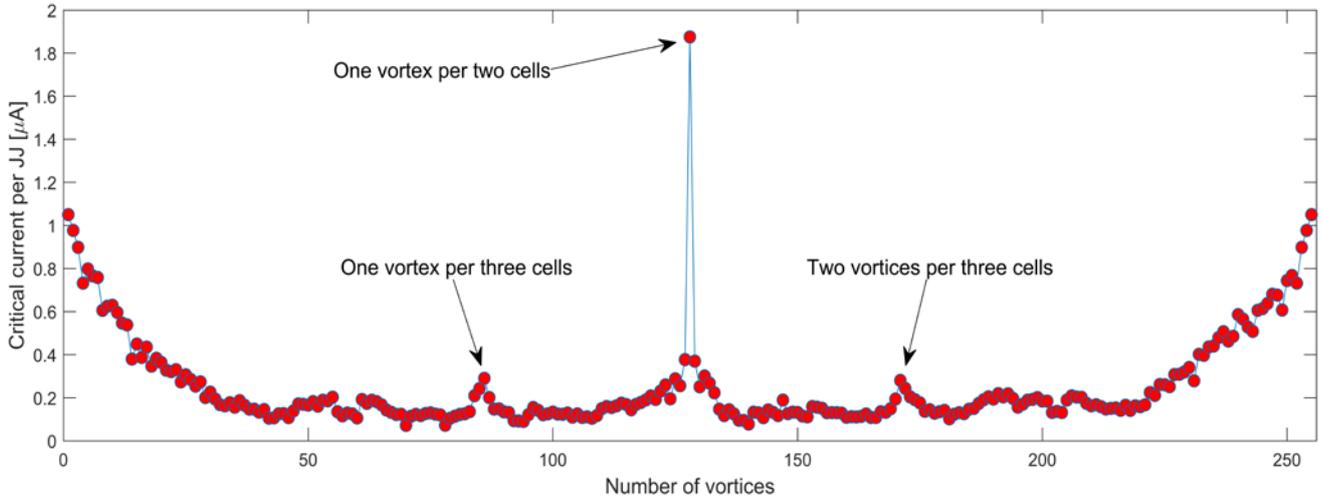

Fig. 3. Normalized (per junction) critical current of the circular long JJ comprized of 256 small junctions as a function of the number of vortices injected in the ring. At 128 vortices the measured critical current corresponds to the theory of two-junction SQUID with half-quantum magnetic bias. Critical currents at zero and 256 vortices are about 25 µA and too large to be shown in the plot scale. Most critical currents are about 0.1 µA. Their value is indirectly defined by a random spread of critical currents of fabricated junctions. There are visible peaks of the critical current at 85, 128, and 170 vortices in the structure. They correspond to matching conditions of 1/3, 1/2, and 2/3 of Josephson vortex per small junction.

## V. CONCLUSION

We have investigated a ring oscillator based on a long circular Josephson junction made of discrete components – 256 small Josephson junctions and inductors. The measured critical current and the energy dissipation in the circuit are about 5× larger than the thermodynamic threshold for applications as a multi-phase clock source in reversible circuits. The circuit served also as a technological benchmark test that allowed us to extract 1σ spreads of the critical currents of comprising junctions, about 0.88 µA or 3.5%, and thereby improve understanding of the effects of fabrication-related spreads on circuit performance. The progress in fabrication and circuit design relative to [5] can be characterized by about a factor of two higher contrasts of critical currents of the ring without and with vortices demonstrated in this work.

## ACKNOWLEDGMENT

We would like to thank Marc Manheimer and Mark Gouker for their interest in and support of this work, and Vlad Bolkhovsky for help with wafer fabrication. V.K. Semenov thanks also M. Cirillo, K.K. Likharev, A.V. Ustinov, and IBM C3 team for informative discussions about dynamics of long Josephson junctions. We thank J. Ren for a unique piece of software developed for her project [5] and reused in this work, and Anatoliy Borodin for help with experimental setup.